\documentclass[aps,prb,twocolumn,showpacs,amsmath,amssymb,floatfix,groupedaddress]{revtex4-1}

\usepackage[utf8]{inputenc}

\usepackage{graphicx}
\usepackage{bm}
\usepackage{color}
\usepackage{epstopdf}
\usepackage{units}
\usepackage{natbib}
\usepackage{ulem}
\begin{document}

\title{Powerful and efficient energy harvester with resonant-tunneling quantum dots}

\author{Andrew N. Jordan$^{1*}$, Bj\"orn Sothmann$^2$, Rafael S\'anchez$^3$ and Markus B\"uttiker$^2$}
\affiliation{$^1$ Department of Physics and Astronomy, University of Rochester, Rochester, New York 14627, USA}
\affiliation{$^2$ D\'epartement de Physique Th\'eorique, Universit\'e de Gen\`eve, CH-1211 Gen\`eve 4, Switzerland}
\affiliation{$^3$ Instituto de Ciencia de Materiales de Madrid (ICMM-CSIC), Cantoblanco, E-28049 Madrid, Spain}

\date{\today}

\newcommand{\mean}[1]{\langle #1 \rangle}           
\newcommand{\cmean}[2]{\,_{#1}\langle #2 \rangle}   

\newcommand{\ket}[1]{|#1\rangle}                    
\newcommand{\bra}[1]{\langle #1|}                   
\newcommand{\ipr}[2]{\langle #1 | #2 \rangle}       
\newcommand{\opr}[2]{\ket{#1}\bra{#2}}              
\newcommand{\pprj}[1]{\opr{#1}{#1}}                 

\newcommand{\Tr}[1]{\mbox{Tr}\left[#1\right]}       
\newcommand{\comm}[2]{\left[#1,\,#2\right]}         
\newcommand{\acomm}[2]{\left\{#1,\,#2\right\}}      
\def\R{\mbox{Re}}                                   
\newcommand{\op}[1]{\hat{#1}}                       
\def\prj{\op{\Pi}}                                  

\newcommand{\oper}[1]{\mathcal{#1}}                 
\newcommand{\prop}[1]{\textit{#1}}                  
\def\gbar{\bar{\gamma}}
\def\ebar{\bar{\eta}}
\def\be{\begin{equation}}
\def\ee{\end{equation}}

\begin{abstract}
We propose a nanoscale heat engine that utilizes the physics of resonant tunneling in quantum dots in order to transfer electrons only at specific energies.  The nanoengine converts heat into electrical current in a multiterminal geometry which permits one to separate current and heat flows.  By putting two quantum dots in series with a hot cavity, electrons that enter one lead are forced to gain a prescribed energy in order to exit the opposite lead, transporting a single electron charge.  This condition yields an ideally efficient heat engine. The energy gain is a property of the composite system rather than of the individual dots.  It is therefore tunable to optimize the power while keeping a much larger level spacing for the individual quantum dots.  Despite the simplicity of the physical model, the optimized rectified current and power is larger than any other candidate nano-engine. The ability to scale the power by putting many such engines into a two-dimensional layered structure gives a paradigmatic system for harvesting thermal energy at the nanoscale.  We demonstrate that the high power and efficiency of the layered structure persists even if the quantum dots exhibit some randomness.
\end{abstract}
\pacs{73.63.-b,85.80.Fi,85.35.-p,84.60.Rb}

\maketitle

\section{Introduction} 
Energy harvesting is the process by which energy is taken from the environment and transformed to provide power for electronics.\cite{white_energy-harvesting_2008} Specifically, thermoelectrics can play a crucial role in future developments of alternative sources of energy. Unfortunately, present thermoelectric engines have low efficiency.\cite{rowe_thermoelectric_2006} Therefore, an important task in condensed matter physics is to find new ways to harvest ambient thermal energy, particularly at the smallest length scales where electronics operate.  Utilizing the physics of mesosopic electron transport for converting heat to electrical power is surprisingly a relatively recent endeavor. While the general relationships between electrical and heat currents and their responses to applied voltages and temperature differences have been understood since the work of Onsager,\cite{onsager_reciprocal_1931} the investigation of thermoelectric properties, and in particular the design of nano-engines has taken the form of several fairly recent concrete proposals. In 1993, Hicks and Dresselhaus investigated the thermoelectric properties of a mesoscopic one-dimensional wire.\cite{hicks_thermoelectric_1993} Mahan and Sofo subsequently showed that the best energy filters are also the best thermoelectrics.\cite{mahan_best_1996} This suggests the use of quantum dots with discrete energy levels to investigate thermodynamic questions.\cite{beenakker_theory_1992,staring_coulomb-blockade_1993} The Seebeck effect - the appearance of a voltage when there is a temperature difference across a sample - was investigated for a single quantum dot with a resonant level by Nakpathomkun {\it et al.}\cite{nakpathomkun_thermoelectric_2010} Resonant levels were also used as energy filters to make a related heat engine from an adiabatically rocked ratchet.\cite{humphrey_reversible_2002} Humphrey {\it et al.} note their model can be generalized to a static, periodic ratchet, which is a quantum version of the model with state-dependent diffusion.\cite{buttiker_transport_1987}  References~\onlinecite{nakpathomkun_thermoelectric_2010,humphrey_reversible_2002} show that a single resonant level is an ideal heat engine, and investigate power and efficiency in that system which has similarities to the present work.

\begin{figure}[t]
\includegraphics[scale=.52]{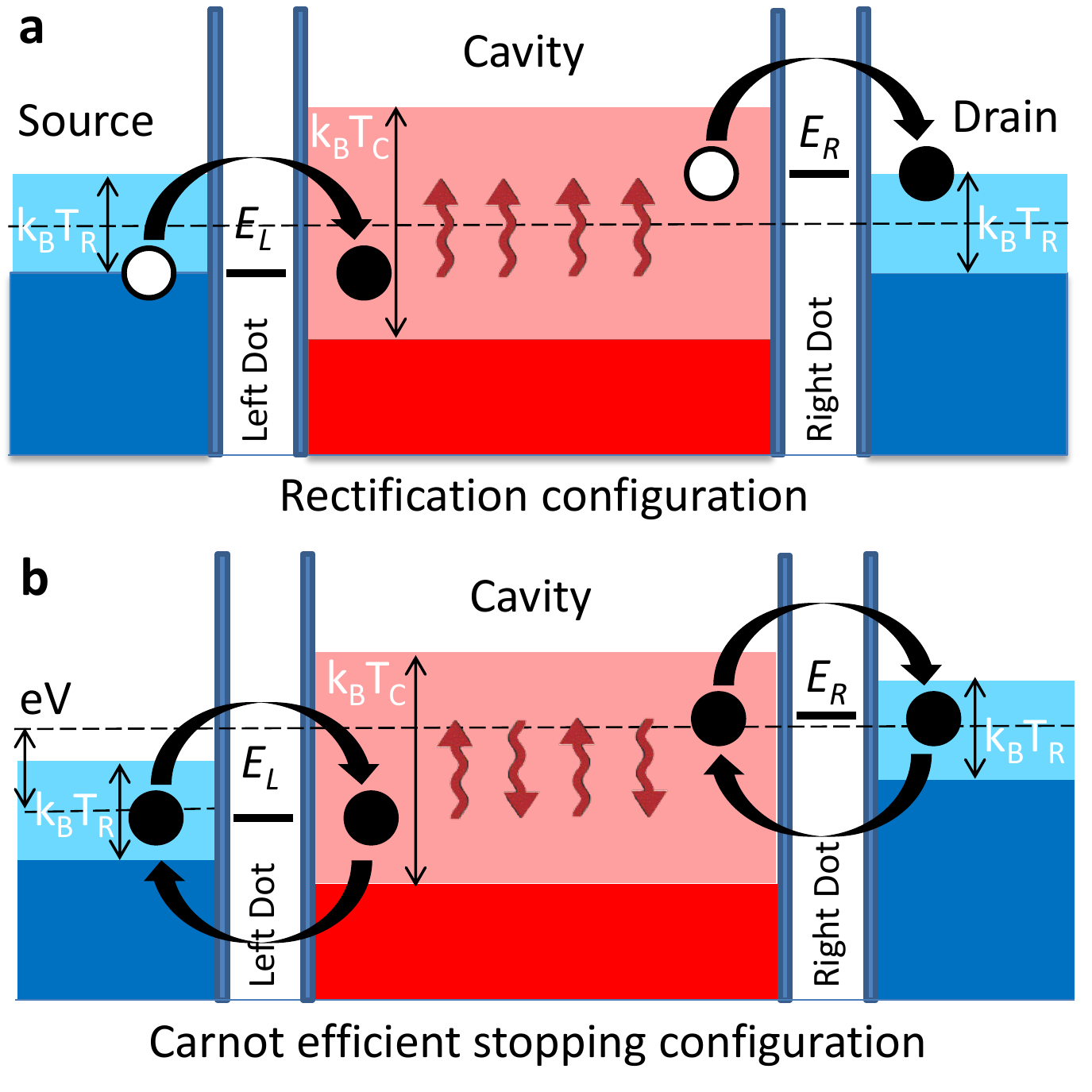}
\caption{Nanoscale heat engine created from a hot cavity connected to cold reservoirs via resonant tunneling quantum dots, each containing a single relevant energy level at energy $E_{L,R}$. The cavity is kept hot through coupling to an energy reservoir at temperature $T_\text{C}$ (not shown) which is considered larger than the reservoir temperature $T_\text{R}$.  Thermal broadening of the Fermi functions in the three regions (source, cavity and drain) is shown by the light shading.  ({a}) Rectification configuration. In the absence of bias (short circuit), electrons enter the cavity via the left lead, gain energy $\Delta E = E_R{-}E_L$ from the cavity, and exit through the right lead, transferring an electrical charge $e$ through the system. ({b}) Carnot efficient stopping (open circuit) configuration. 
} \label{fig1}
\end{figure}

Coulomb-blockaded dots can be ideally efficient converters of heat to work, both in the two-terminal \cite{esposito_thermoelectric_2009} and three-terminal\cite{sanchez_optimal_2011} cases; however, since transport occurs through multiple tunneling processes, the net current and power are very small.  
In light of the small currents and power produced by Coulomb-blockaded quantum dots, open cavities with large transmission that weakly changes with incident electron energy have been considered.\cite{sothmann_rectification_2012} While this system produces more rectified current than Coulomb-blockaded quantum dots, simply increasing the number of quantum channels does not help because the energy dependence of transmissions in typical mesoscopic conductors is a single-channel effect even for a many-channel conductor. Hence, the rectified current and power drop as channel number is increased unless special engineering of the contacts is made. Consequently, we should optimize the rectified current and electrical power for a strongly non-linear system, operating in the single channel limit. We do this here with the physics of resonant tunneling through barriers connected to the hot energy source.  We show this achieves the maximum current a single channel can give, which we conjecture is the optimal configuration.  Importantly, in our model the energy source is separate from the electrical circuit, as is required for energy harvesting. Therefore no charge is extracted from it and the optimal configuration is independent of the parameters of the energy source.  This is different from single quantum dot proposals \cite{nakpathomkun_thermoelectric_2010,humphrey_reversible_2002,esposito_thermoelectric_2009} where the leads are held at different temperatures, and the amount of transferred heat depends on the lead chemical potentials.

Resonant tunneling is a quantum mechanical effect, where constructive interference permits an electron tunneling through two barriers to have unit transmission. This is only true if the electron has a particular energy equal to the bound state in the quantum dot, or within a range of surrounding energies, whose width is the inverse lifetime of the resonant state.\cite{buttiker_coherent_1988}  Electrons with any other energy are effectively forbidden from transmitting across the quantum dot. 
In this way, a resonant tunneling barrier acts like an energy filter: only those electrons that match the resonant condition are permitted to pass.   We assume for simplicity the resonant tunnel barrier (or the dot) is symmetrically coupled.

The geometry we consider is related to one that has already been fabricated experimentally, but considered for a different purpose, based on the theoretical proposal of Edwards {\it et al.}.\cite{edwards_quantum-dot_1993,edwards_cryogenic_1995} Prance {\it et al.} performed experiments on this system in its dual role as an electronic refrigerator.\cite{prance_electronic_2009}  They demonstrated that applying bias to the system results in cooling a large \unit[6]{$\mu$m$^2$} cavity from \unit[280]{mK} to below \unit[190]{mK}.
We are primarily interested in cavities made as small as possible while still having good thermal contact with the heat source. This further miniaturization permits many engines to be put in parallel and give a large output power.

The paper is organized as follows. In Sec. \ref{model} we describe the transport through our device, the results of analytical and numerical calculations are presented in Sec. \ref{results}. Section \ref{layered} investigates the scaling of the simple system consisting of two quantum dots up to a layered structure with many channels contributing at the same energies and discusses the robustness of the proposed structure to fluctuations in the fabrication process. The conclusions are presented in Sec. \ref{conclusions}.

\section{Model} 
\label{model}

The model we consider (shown in Fig.~1) consists of a cavity connected to two quantum dots, each with a resonant level of width $\gamma$ and energy $E_{L,R}$. We consider the situation where the widths are equal, while the energy levels are different (these are controlled by gate voltages).  The energy difference $\Delta E = E_{R} - E_{L}$  is an important energy scale of our composite system, which we refer to as the {\it energy gain}. It is distinct from the level spacing $\delta$ in the individual dots as well as from the level width.  The nano-cavity the dots are connected to is considered to be in equilibrium with a heat reservoir of temperature $T_\text{C}$ that is hotter than the left and right electron reservoirs, having chemical potentials $\mu_{L,R}$ and equal temperatures, $T_\text{R}$. We assume strong electron-electron and electron-phonon interactions relax the electron energies as they enter and leave the cavity, so the cavity's occupation function may be described with a Fermi function, $f(E- \mu, T)=1/(1+\exp[(E-\mu)/k_B T])$ completely characterized by a cavity chemical potential $\mu_\text{C}$ and temperature $T_\text{C}$, with $k_\text{B}$ being the Boltzmann constant. This process of inelastic energy mixing is assumed to occur on a faster time scale than the dwell time of an electron in the cavity. Thermal energy flows from the coupled hot bath into the cavity as a heat current, and keeps the temperature different from that of the electron reservoirs. The nature of the heat reservoir is not specified in this model, but refers quite generically to any heat source we wish to harvest energy from.  Our setup should be contrasted with the more widely studied two-terminal configurations in which electric and heat currents flow in parallel. Our model permits a separate heat circuit to the central reservoir and a separate (transverse) electrical circuit.

The chemical potential of the cavity and its temperature (or equivalently, the incoming heat current) are constrained by conservation of global charge and energy.   These constraints are given by the simple equations, $I_L {+} I_R{=}0$, and $J_L {+} J_R {+} J{=}0$ in the steady state, where $I_{L,R}$ is electrical current in the left or right contact, and $J_{L,R}$ the energy current.  Energy current is seemingly not conserved because of the heat current $J$ flowing from the hot reservoir. 

The currents $I_j$, $j{=}L,R$, are given by the well known formulas $I_j {=} (2e/h) \int dE\, T_j(E) [f_j {-} f_C]$ and $J_j = (2/h) \int dE\, T_j(E) E [f_j  {-} f_C]$, where $T_j(E)$ is the transmission function of each contact for each incident electron energy $E$.  In our quantum dot geometry, the resonant levels give rise to a transmission function of Lorentzian shape,\cite{buttiker_coherent_1988} 
\be
T_j(E){=} \frac{\Gamma_1 \Gamma_2}{(E{-}E_j)^2 {+} \left(\frac{\Gamma_1+\Gamma_2}{2}\right)^2},
\ee 
where $\Gamma_{1,2}$ are the attempt frequencies of the two barriers of the resonant quantum dot(s). Here we assume symmetric coupling for simplicity, $\Gamma_1 = \Gamma_2 = \gamma$, so $\gamma$ is the width of the level, or inverse lifetime of an electron in the dot.  Note that the Lorentzian energy dependence applies if the level width $\gamma$ is small compared to the level spacing $\delta$ in the individual quantum dots. A crucial advantage of our setup which we will exploit later on is that it permits the use of small dots with a level spacing that is large compared to temperature but with the energy gain and level width of each dot on the order of the temperature.

In the limit where the width of the level is smaller than the thermal energy in the cavity/dot system, $\gamma {\ll} k_B T_\text{C}, k_B T_\text{R}$, the transmission will pick out only the energies $E_L$ or $E_R$ in the above energy integral expressions for the currents giving simple equations.   Consequently, we have these equations for the conservation laws for charge and energy:
\begin{subequations}
\begin{eqnarray}
0 &=& f_L - f_{\text{C}L} +  f_R - f_{\text{C}R}, \label{conservation1}\\
0 &=& J h/(2 \gamma)+E_L [ f_L - f_{\text{C}L}] + E_R [ f_R - f_{\text{C}R}], 
\label{conservation2}
\end{eqnarray}
\end{subequations}
where $f_L =  f(E_L {-} \mu_L, T_\text{R})$, $f_R =  f(E_R {-} \mu_R, T_\text{R})$, $f_{\text{C}L} =  f(E_L {-} \mu_\text{C}, T_\text{C})$, and $f_{\text{C}R} =  f(E_R {-} \mu_\text{C}, T_\text{C})$.
From these two equations, we can solve for (say) the quantity $f_{\text{C}R}-f_R=Jh/(2\gamma\Delta E)$.
This quantity is proportional to the electrical current through the left lead $I_L = -I_R \equiv I$, the net current flowing through the system. 

A solution of Eqs. \eqref{conservation1} and \eqref{conservation2} to linear order in the deviation of the cavity's temperature and chemical potential from the electronic reservoirs indicates that the maximal power of the heat engine will be produced when the chemical potentials of the reservoirs are symmetrically placed in relation to the average of the resonant levels, $\mu_{R,L}{=}{\pm}\mu/2 {+} (E_L {+} E_R)/2$.  For this special case, an exact solution is possible because the constant solution $\mu_\text{C} {=} (E_L {+} E_R)/2$ for the cavity chemical potential satisfies the charge conservation condition \eqref{conservation1} for all temperatures.  

\section{Results}
\label{results}
\subsection{Limit of small level width}
We now describe the physics of this nano-engine.  We first focus on the regime $\gamma\ll k_\text{B}T_\text{R},k_\text{B}T_\text{C}$, which can be analyzed analytically and will afterwards discuss the regime $\gamma{\sim} k_\text{B}T_\text{R},k_\text{B}T_\text{C}$ which we numerically find to yield the largest current and power. Physically, if an electron comes in the left lead at energy $E_L$ and exits the right lead with energy $E_R {>} E_L$, it must gain precisely that energy difference $\Delta E = E_R - E_L$.  Thus, in the steady state, any incoming heat current $J$ must be associated with an electrical current $I$, with a conversion factor of the energy gain, $\Delta E$, to the quantum of charge, $e$,
\be
I = \frac{e J}{\Delta E}.
\ee
This results holds regardless of what bias is applied or what the temperature is. 

The efficiency of our heat engine, $\eta$, is defined as the ratio of the harvested electrical power $P{=} | (\mu_L {-} \mu_R) I|/e$ to the heat current from the hot reservoir, $J$.
For our system it takes a particularly simple form,
\be
\eta = \frac{|\mu_L - \mu_R|}{\Delta E}.
\label{efficiency}
\ee

In order to proceed further, we must find the chemical potential of the cavity and its temperature given in terms of the incoming heat current and chemical potentials and temperature of the electron reservoirs. These are found by employing the principle of conservation of global charge and energy, see Eqs.~\eqref{conservation1} and \eqref{conservation2}.  

\begin{figure}[tb]
\includegraphics[scale=.28]{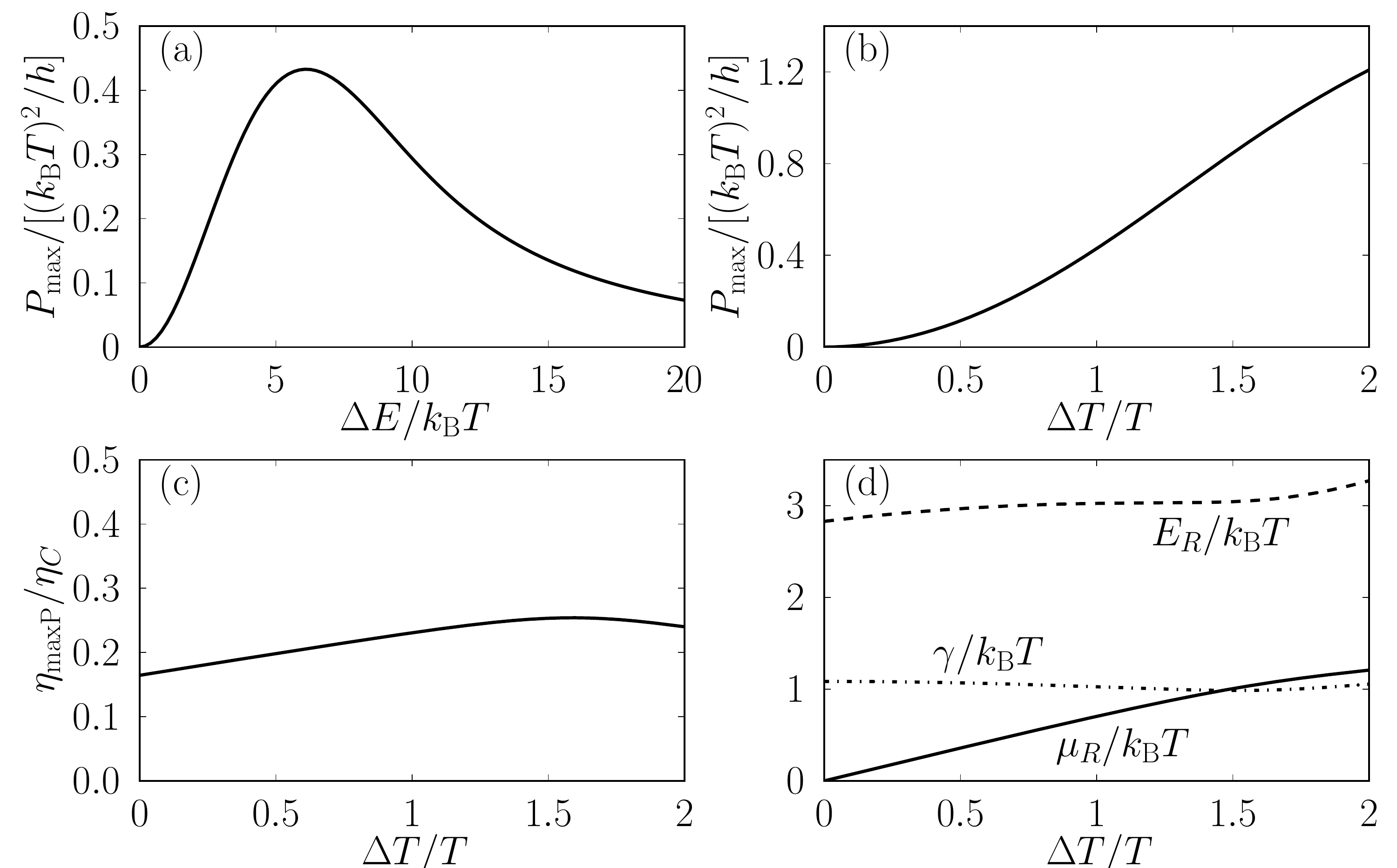}
\caption{({a}) Scaled maximum power as a function of energy gain $\Delta E$ for $\Delta T = T$ and  level width $\gamma$ and $\mu_R$ optimized to give maximum power.  ({b}) Scaled maximum power as a function of $\Delta T/T$ for optimized values of $E_R = \Delta E/2 , \gamma$ and $\mu_R$.    ({c})  Efficiency at maximum power for the values of $E_R = \Delta E/2, \gamma$ and $\mu_R$ chosen to maximize power.   ({d}) The optimized values are plotted versus $\Delta T/T$. Here, $E_R = \Delta E/2$, $\mu_R =\mu/2$.} \label{fig2}
\end{figure}

Rather than considering the cavity temperature $T_\text{C}$ as a function of the heat current $J$ we are harvesting, we can turn our perspective around, and keep the cavity temperature $T_\text{C}$ fixed from being in thermal equilibrium with the hot energy source.  With this insight, we can express the heat current $J$ in terms of the hot cavity temperature and other system parameters.  We find in the limit where $\gamma \ll k_B T_\text{R}, k_B T_\text{C}, \Delta E$,
\be
J = \frac{2\gamma \Delta E}{h} [f(\Delta E/2, T_\text{C}) - f(\Delta E/2 - \mu/2, T_\text{R})],
\label{jsol}
\ee
which satisfies charge and energy current conservation.  In Eq.~(\ref{jsol}), $h$ is Planck's constant.
  
Importantly, without bias there is a rectified electrical current given by 
\be
I = e J/\Delta E \approx \frac{e \gamma \Delta E}{4 h}[(k_B T_\text{R})^{-1} - (k_B T_\text{C})^{-1}],
\label{isol}
\ee 
in the limit where $k_B T_\text{R}, k_B T_\text{C} \gg \Delta E$.   This current is driven solely by the fixed temperature difference between the systems.   We note both the heat and electrical current are proportional to $\gamma$, the energy width of the resonant level.  Consequently, the currents and power produced in this system will tend to be small since we have assumed that $\gamma$ is the smallest energy scale.  It is also clear that both are controlled by the size of $\Delta E$, so increasing this energy gain will improve power until it exceeds the temperature.  Later we will generalize these results by numerically optimizing the power produced in this nano-engine.

In order to harvest power from this rectifier, a load should be placed across it. Equivalently, we could apply a bias $V=\mu/e$ to this system tending to reduce the rectified current.   At a particular value, $\mu_{\rm stop}$, the rectified current vanishes, giving the maximum load or voltage one could apply to extract electrical power at fixed temperatures $T_\text{R}, T_\text{C}$.  This value is found when $J$ and $I$  vanish, given by Eq.\eqref{jsol}:
\be
\mu_{\rm stop} = \Delta E  \left(1 - \frac{T_\text{R}}{T_\text{C}}\right).
\label{stop}
\ee
Consequently, the voltage applied must not be larger than $\mu_{\rm stop}/e$ and therefore from Eq.~(\ref{efficiency}), the efficiency is bounded by $\eta \le 1 - \frac{T_\text{R}}{T_\text{C}} = \eta_\text{C}$.
At the stopping voltage, the thermodynamic efficiency attains its theoretical maximum, the Carnot efficiency, $\eta_\text{C}$, showing this system is an ideal nanoscale heat engine.   Naturally, at this point [see Fig.~\ref{fig1}(b)] the system is reversible with no entropy production.
 Also interesting is the efficiency at the bias point where power is maximum.  For temperature larger than $\Delta E$ or $eV {=} \mu$, we can approximate the Fermi functions to find $P \approx (\gamma / 4 h k_B T_\text{R}) \mu (\mu_{\rm stop} {-} \mu)$, resulting in a parabola as a function of $\mu$, with maximum power 
\be
P_{\rm max} \approx  \frac{\gamma \Delta E^2 \eta_C^2}{16 h k_B T_R},
\label{pmax}
\ee
and efficiency $\eta_\text{maxP} = \eta_\text{C}/2$ which is in agreement with general thermodynamic bounds for systems with time-reversal symmetry.\cite{van_den_broeck_thermodynamic_2005,esposito_universality_2009,benenti_thermodynamic_2011}

\subsection{Optimization}

One can go beyond this limit for the efficiency by solving the conservation laws numerically.  We optimize the total power produced by the heat engine by varying the resonance width $\gamma$, as well as the energy gain $\Delta E$ and applied bias $V {=} \mu/e$, given fixed temperatures $T_\text{R}, T_\text{C}$.  These results are plotted in Fig.~\ref{fig2}, where we define the average temperature $T {=} (T_\text{C} {+} T_\text{R})/2$, and its difference, $\Delta T {=} T_\text{C} {-} T_\text{R}$.  In Fig.~2(a), we see that for $\Delta E < k_B T$ the power increases as $\Delta E^2$, as indicated in Eq.~(\ref{pmax}), but then levels off and decays exponentially, attaining its maximum around $\Delta E = 6 k_B T$.  Similarly, the choice $\gamma = k_B T$ gives optimal power.   We emphasize that $\Delta E$ and the level width are two essentially independent energy scales in our system. As a matter of fact the energy gain $\Delta E \approx 6 k_B T$ is almost an order of magnitude larger than the level width $\gamma \approx k_B T$. These considerations suggest an experimental strategy for maximizing the power of such a device: Measure what the resonant level widths are, and tune the reservoir temperatures and energy gain (with the help of gate voltages), to it. 

From Fig.~\ref{fig2}(b) we also see that the efficiency at maximum power drops from half the Carnot efficiency to about 0.2$\eta_\text{C}$ when the parameters are optimized.  However, we note that when $\gamma$ is kept small, in the nonlinear regime the efficiency can  exceed the bound $\eta_\text{maxP}\leq\eta_\text{C}/2$ found in the linear regime. This small drop in efficiency is more than compensated by the extra power we obtain.  
%
%
Importantly, according to Fig.~\ref{fig2}, the power reaches a maximum of $P_{\rm max} {\sim} 0.4 (k_B \Delta T)^2/h$, or about \unit[0.1]{pW} at $\Delta T$=\unit[1]{K}, a two order of magnitude increase from a weakly nonlinear cavity. \cite{sothmann_rectification_2012} This jump in power can be attributed to the highly efficient conversion of thermal energy into electrical energy by optimizing both the level width and energy level difference.  
Compared to a heat engine based on resonant tunneling through a single quantum dot in a strictly two-terminal geometry,\cite{nakpathomkun_thermoelectric_2010} our three-terminal energy harvester gains a factor of 2 in power while achieving the same efficiency.
To give some perspective on this output, if a \unit[1]{cm$^2$} square array of these nanoengines were fabricated, each occupying an area of \unit[100]{nm$^2$}, they would produce a power of \unit[0.1]{W}, operating at $\Delta T$ = \unit[1]{K}.

\begin{figure}[tbp]
\includegraphics[width=\columnwidth]{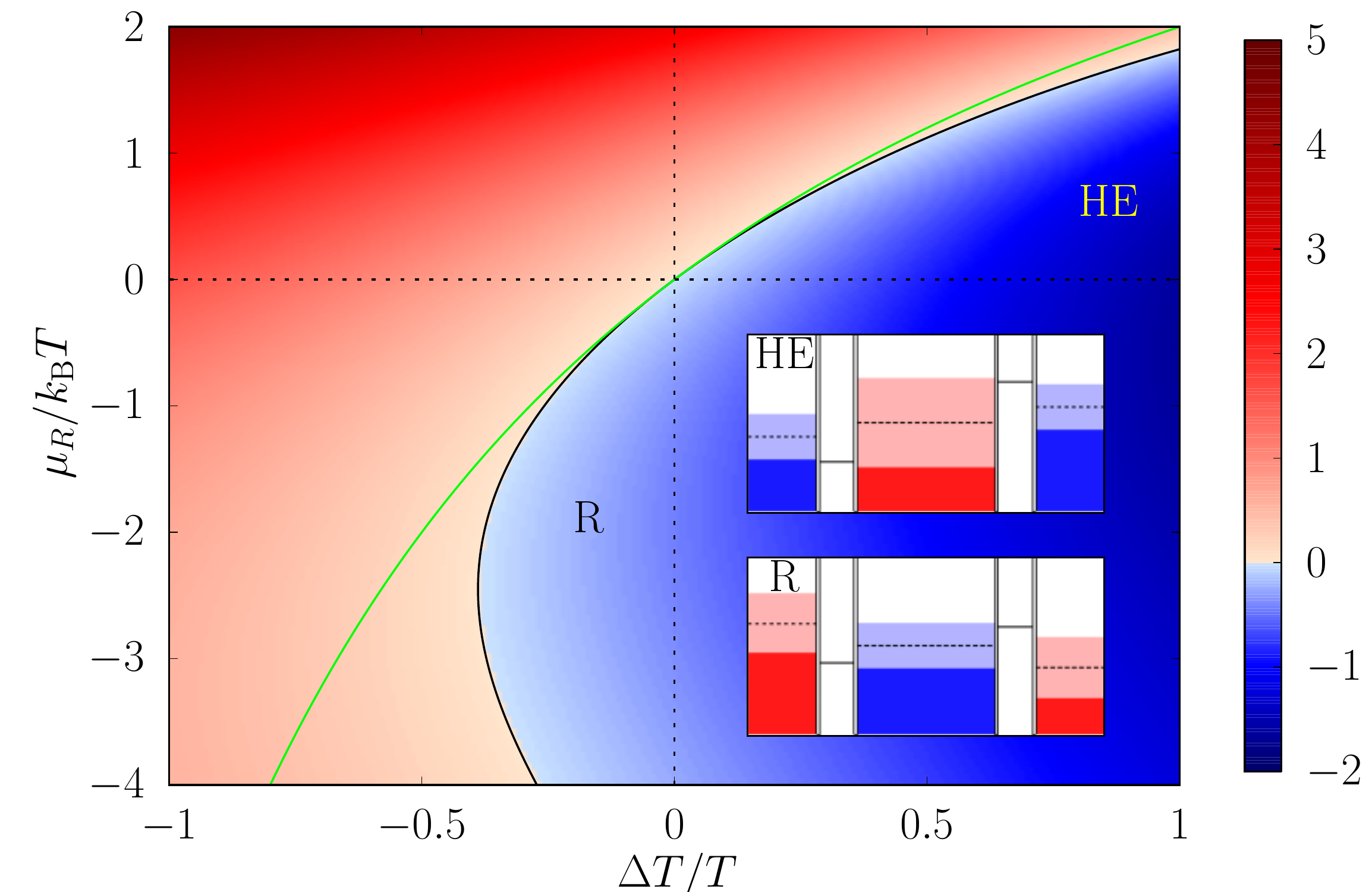}
\caption{Scaled heat current $J/[2(k_\text{B}T)^2/h]$ leaving (blue) or entering (red) the central cavity vs temperature difference (x axis) and applied bias (y axis) divided by average temperature.  The plot is for system parameters optimized for maximal power output,  an energy gain of $\Delta E\approx6 k_B T$ and level width $\gamma\approx k_B T$. The green line is the $J=0$ curve for $\gamma \rightarrow 0$ while the black curve is the $J=0$ line for the optimized $\gamma$.  The system can work as a heat engine (HE) in the blue region for $\Delta T>0, \mu_R = \mu/2 > 0$ (configuration shown in the inset labeled HE), or as a refrigerator (R) in the blue region for $\Delta T <0, \mu_R = \mu/2 < 0$ (configuration shown in the inset labeled R). \label{heatcurr}}
\end{figure}

In Fig.~\ref{heatcurr}, the heat current $J$ is plotted versus temperature difference and applied bias.  There we find that when system parameters are optimized to give maximum power, the system can be operated in the mode of a heat engine (HE) or a refrigerator (R). However, in contrast to the case where the levels are narrow compared to the other energy scales [and consequently the cavity can cool to arbitrarily low temperatures in principle; see green solid line, Eq.~\eqref{stop}], for this choice of parameters the cavity will only cool to the temperature where the $J=0$ curve (solid black line) bends back.

\begin{figure}[tbp]
\includegraphics[width=0.8\columnwidth]{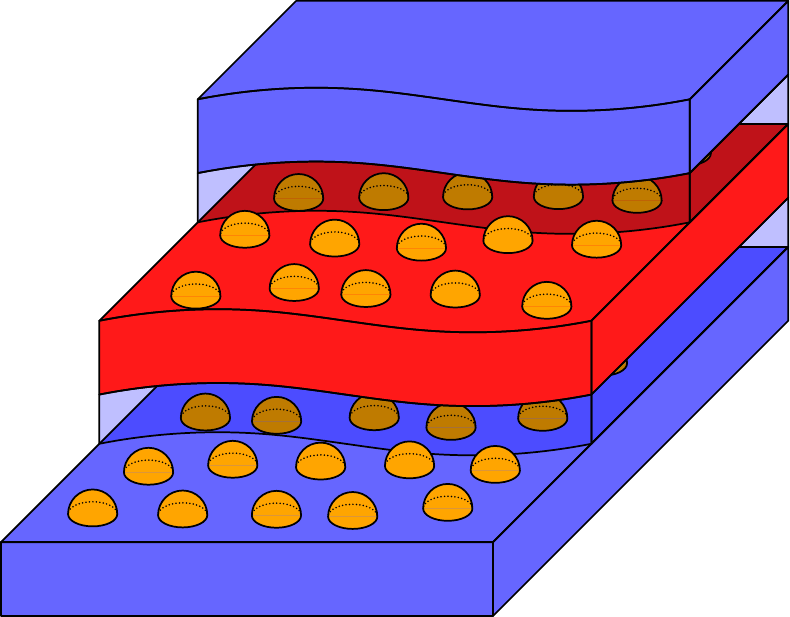}
\caption{Self-assembled dot engine: A cold bottom electrode (dark blue) is covered with a layer of quantum dots (orange) embedded in an insulating matrix (transparent blue). The quantum dot layer is covered by the hot central region shown in dark red. On top of it, there is another quantum dot layer. The whole structure is terminated by a cold top electrode (dark blue). Importantly, the positions of the quantum dots in the two layers do not have to match each other. Thus, the device can be realized using self-assembled quantum dots. The latter can have charging energies and single-particle level spacings of the order of \unit[10]{meV} (Refs. \onlinecite{drexler_spectroscopy_1994,jung_shell_2005}), thereby allowing the nanoengine to operate at room temperature.
\label{swiss}}
\end{figure}

\subsection{Scaling} 

In reality, there will be other quantum dot resonant levels the electron can occupy that are higher up in energy.  We have assumed that the cavity temperature and applied bias are sufficiently small so transport through these levels can be neglected.  Our model is quite general, so it may be applied to both semiconductor dots in two-dimensional electron gases,\cite{scheibner_thermopower_2005} as well as three dimensional metallic dots.\cite{averin_resonant_1997}  This latter case is quite interesting since one can fabricate an entire plane of repeated nanoengines in parallel in order to scale the power.\cite{park_formation_2004}  Furthering this idea, for such a repeated array of cavities and quantum dots, one can connect all the cavities to make a single engine, see Fig.~\ref{swiss}.  The two boundary layers consist of planes of quantum dots, so electrons can only penetrate through them.  These layers sandwich a hot interior region and separate it from the left and right cold exterior contacts.  This is equivalent to taking a large cavity with two leads and scaling the power by adding more quantum dots (rather than trying to add more channels to a single contact).
Interestingly, the layered structure can help to reduce phononic leakage heat currents that would otherwise reduce the efficiency of the system: Phonons scatter efficiently at interfaces, and the random dot arrangement will further reduce phonon related heat losses.  The sandwich engine fabricated with self-assembled quantum dots tolerates variations in width and fluctuations in energy levels, as we will see in the next section.

\section{Robustness of the layered quantum dot engine to fabrication fluctuations}
\label{layered}
\begin{figure}[t]
\includegraphics[width=\columnwidth]{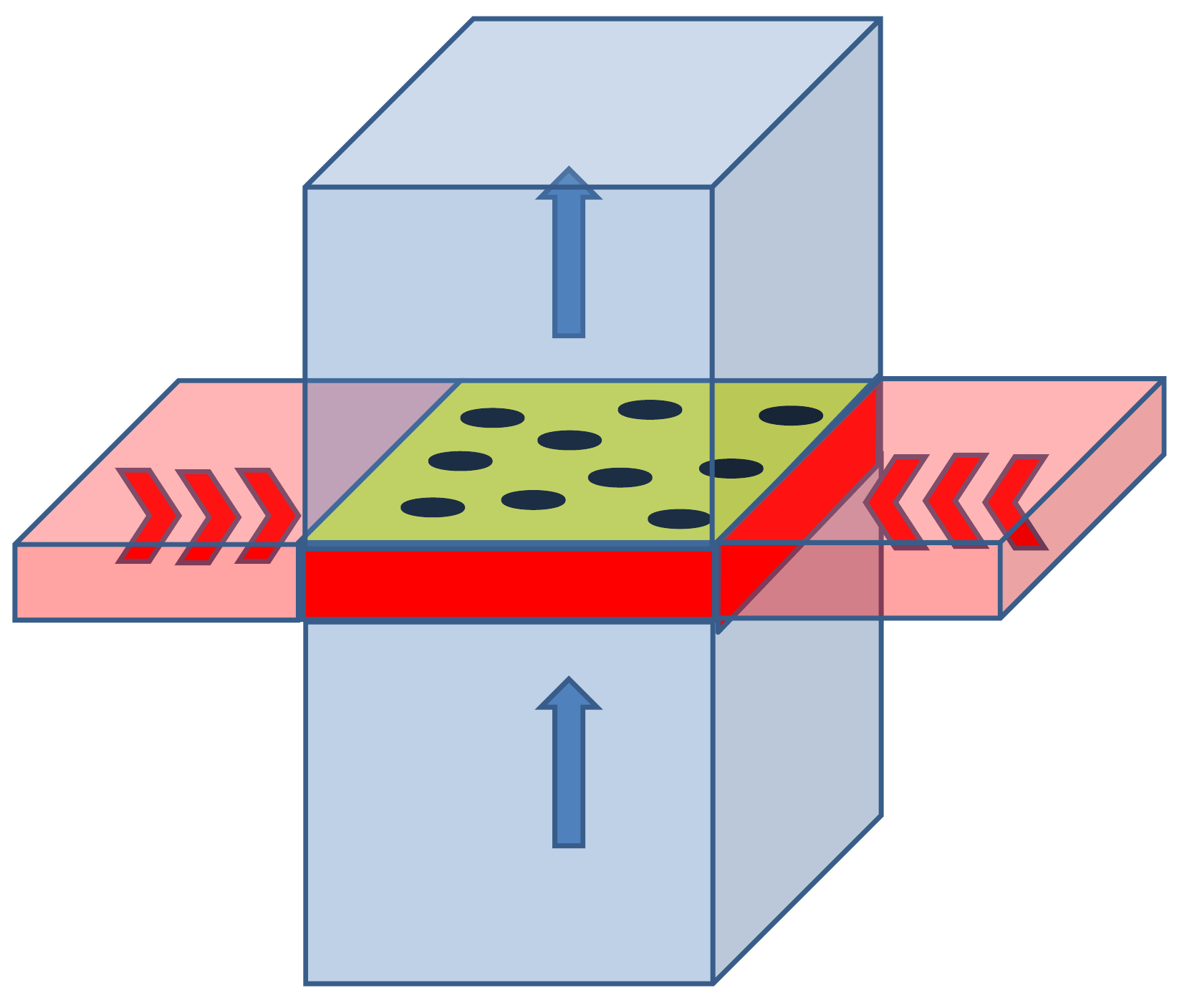}
\caption{Schematic of the system operation configuration. Heat enters the engine from the pink hot region, indicated by the red chevrons.  The engine itself is signified by the yellow sandwich with black holes indicating the position of the resonant tunneling quantum dots.  The position of the holes can be disordered, similar to a slice of Swiss cheese. Electrical current is generated perpendicular to the layers, indicated with the blue arrows flowing in the light blue cold region.}
\label{swisscheese}
\end{figure}

Here, we further investigate the self assembled quantum dot layered engine, and what some of its theoretical characteristics are under realistic fabrication conditions. The basic operating configuration for the engine is shown in Fig.~\ref{swisscheese}.  Heat flows from the hot energy source we wish to harvest energy from into the engine, and is converted into electrical power, along with residual thermal energy, dumped into the cold temperature bath. The electrical current is carried by the cold thermal bath, where it powers a load and then completes the electrical circuit on the opposite cold terminal.  The flow of heat out of the hot energy source will consequently tend to cool the hot source. The energy harvesting application we primarily have in mind is taking heat away from computer chips and running other devices on the chip itself.  In electrical chips, thermal energy is an abundant and free resource.  Indeed, heat is not only free, it is a nuisance preventing further improvements on chip technology. The fact that the proposed heat engine not only harvests the thermal energy, converting some of it into electrical power, but also cools the hot source is therefore a side benefit to the proposal of including nanoscale heat engines as part of an emerging chip technology.

In the actual fabrication of a resonant tunneling nano-engine, as long as there are only a few dots, the precise placement of the resonant levels can be controlled by gate voltages in order to maximize the power generated by the engine. However, as soon as we consider self-assembled quantum dots with charging energies and single-particle level spacings of the order of \unit[10]{meV},\cite{drexler_spectroscopy_1994,jung_shell_2005} thereby allowing the nano-engine to operate at room temperature, this kind of control is out of the question.  To make such an engine, there are several possible fabrication techniques that could be employed using layers of quantum dots and wells to have the resonant energy levels lower than the Fermi energy on one side of the heat source, and higher on the other side.
However, in all of these fabrication methods, the growth of quantum dots does not occur at a perfectly regular rate, so it is natural to expect there will be variation of the resonant energy level from dot to dot.  We must then check whether this fact will degrade the performance of the engine, and if so by how much.

To answer this question, we consider the energy-resolved current arising from a number of electrons passing through $N$ quantum dots on the left and then the right layer.  The total current coming from the left slice is given by
\begin{eqnarray}
I_L^{\rm tot} &=& \frac{2e}{h} \int dE T_{{\rm eff}, L} [f(E - \mu_L, T_{\rm R}) - f(E - \mu_{\rm C}, T_{\rm C})], \nonumber \\
T_{{\rm eff}, L} &=&  \sum_{i=1}^N T_i(E, E_i),
\end{eqnarray}
with similar equations for the total right current, as well as the energy currents.     Here, $T_i(E, E_i)$ is the transmission probability of quantum dot $i$, which has a resonant energy level $E_i$ and a width $\gamma_i$, $T_i(E, E_i) = \gamma_i^2/[(E-E_i)^2 + \gamma_i^2]$ for symmetrically coupled quantum dots.   Since neither the left nor cavity Fermi functions depend on the level placement, the sum over the quantum dots can be done to give an effective transmission function $T_{{\rm eff},L}$ for the whole left slice.
We can make further progress by assuming the fabrication process can be described as a Gaussian random one, where the energy level $E_i$ is a random variable with an average of $E_L$ and a standard deviation of $\sigma$.  For simplicity, we only consider random variation in $E_i$, but there will also be variation in $\gamma_i$ we ignore for the present.  With this model, the effective transmission will have the average value
\begin{figure}[t]
\includegraphics[width=\columnwidth]{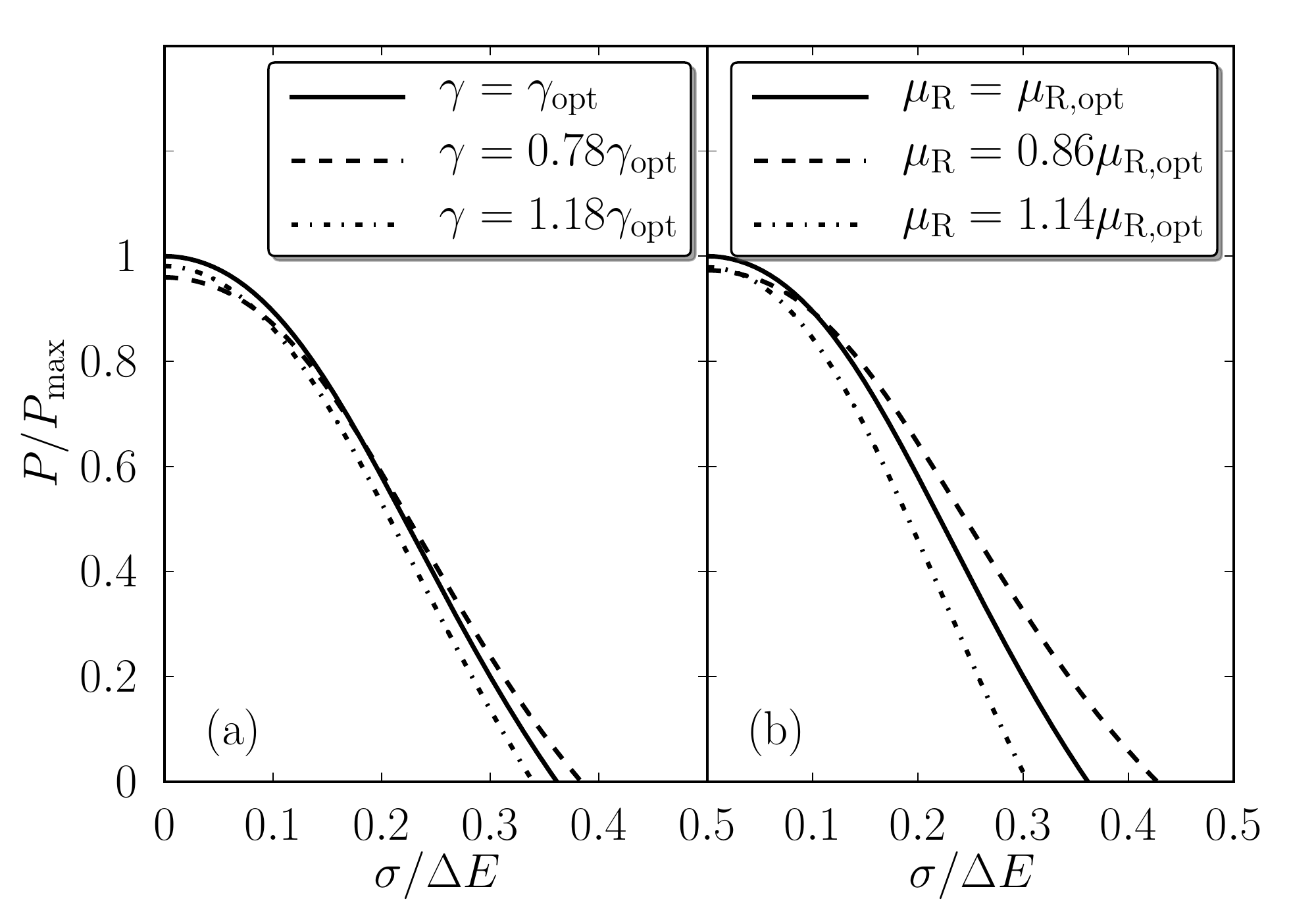}
\caption{The power per nanoengine is plotted vs the width of Gaussian random distribution of energy levels.  The power is normalized to the maximum power a single nanoengine can give (for optimized parameters), while the distribution width is plotted in units of $\Delta E$, the average energy gain between left and right dots.  (a) The power is plotted for different level widths of the dots, where $\gamma_{\rm opt} = 1.02\, k_B T$.    (b) The power is plotted for different applied voltages, where $\mu_{R,{\rm opt}} = 0.7\, k_B T$. }
\label{ensemblepower}
\end{figure}

\be
\langle T_{{\rm eff},L} \rangle = N \langle T_i(E, {\bar E}) \rangle_P = N \int d{\bar E}\, T(E - {\bar E}) P_G({\bar E}, \sigma),
\ee 
where $P_G$ is the Gaussian distribution described above. Thus, we see the effective transmission function is simply a convolution of the Lorentzian transmission function and the Gaussian distribution, known as a Voigt profile.  This leads to further broadening of the Lorentzian width.  With these considerations, the conservation laws for charge and energy retain the same basic form as described in Eqs.~(\ref{conservation1}) and (\ref{conservation2}), but with $N$ times the Voigt profile playing the role of the energy-dependent transmission for the left and right leads.  We have numerically solved these equations and plotted the maximum power per nanoengine versus the width of the Gaussian distribution in Fig.~\ref{ensemblepower}.    The parameters are chosen so as to optimize the engine's performance without any randomness in the level position.  As the randomness of the level position is increased, the power begins to drop as expected.  However, even when the scatter of the energy levels is $10\%$ of $\Delta E$, the power only drops to $90\%$ of its maximum, showing that this engine is robust to these kinds of fluctuations in fabrication.  
Notice that if the level width is less than the optimal amount, some disorder in the level energy can actually {\it improve} performance in comparison to a cleanly optimized level width.  This is because of the additional broadening in energy space the level disorder provides.   
Another interesting effect is shown in Fig.~\ref{contourpower}, where the value of $\gamma$ is not optimal together with a given amount of disorder in the energy level positions.  The figure demonstrates that even in this experimentally realistic case, a change of the voltage and relative level spacing can yield results which are nearly as good as the optimal case.
\begin{figure}[t]
\includegraphics[width=\columnwidth]{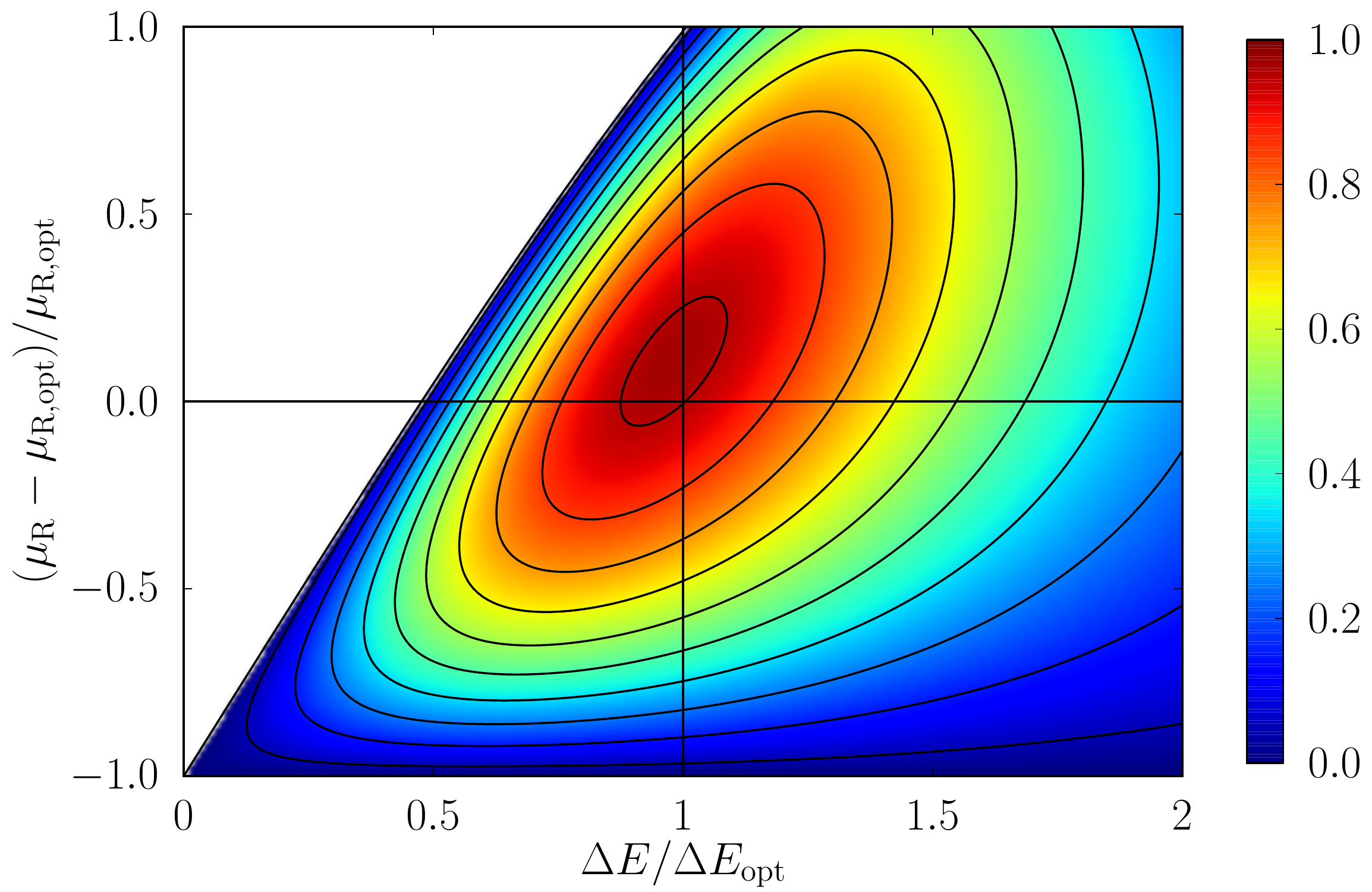}
\caption{The power per nanoengine is plotted vs the right chemical potential and the average energy gain, $\Delta E$, in units of the optimal power, $P_{\rm max}$.  The level width is taken to be a suboptimal value, $\gamma = 0.8 k_B T$ and the width of Gaussian random distribution is fixed at $\sigma = 0.2 k_B T$.  The subscript ``opt" refers to the parameter that is optimized for maximal power in the clean case.}
\label{contourpower}
\end{figure}

\section{Conclusions}
\label{conclusions}
In this paper, we have opened a route to highly efficient solid state energy harvesting.  Our work shows that the most efficient structure is found with energy filters that have transmission probabilities close to 1 in a configuration where the level structure of spatially separated quantum dots can be independently adjusted to give the desired energy gain. This yields a high power heat engine with no moving parts.  We show that nano-scale quantum dots can be employed in a parallel configuration that delivers substantial power. 

This work was supported by the US NSF Grant No. DMR-0844899, the Swiss NSF, the NCCR MaNEP and QSIT, the European STREP project Nanopower, the CSIC and FSE JAE-Doc program, the Spanish MAT2011-24331 and the ITN Grant No. 234970 (EU).  A.N.J. thanks M.B. and the University of Geneva for hospitality, where this work was carried out.

$^*$Corresponding author: jordan@pas.rochester.edu
\vspace{-.5cm}


\begin{thebibliography}{25}

\expandafter\ifx\csname natexlab\endcsname\relax\def\natexlab#1{#1}\fi
\expandafter\ifx\csname url\endcsname\relax
  \def\url#1{\texttt{#1}}\fi
\expandafter\ifx\csname urlprefix\endcsname\relax\def\urlprefix{URL }\fi


\bibitem[{White(2008)}]{white_energy-harvesting_2008}
B.~E. White, 
\newblock {Nature Nanotech.} \textbf{3}, 71 (2008).

\bibitem[{Rowe(2006)}]{rowe_thermoelectric_2006}
D.~M. Rowe, 
\newblock {I. J. Innovations Energy Syst. Power} \textbf{1}, 13 (2006).

\bibitem[{Onsager(1931)}]{onsager_reciprocal_1931}
L. Onsager, 
\newblock {Phys. Rev.} \textbf{37}, 405 (1931).

\bibitem[{Hicks \& Dresselhaus(1993)}]{hicks_thermoelectric_1993}
L.~D. Hicks and M.~S. Dresselhaus, 
\newblock {Phys. Rev. B} \textbf{47}, 16631 (1993).

\bibitem[{Mahan \& Sofo(1996)}]{mahan_best_1996}
G.~D. Mahan and J.~O. Sofo, 
\newblock {Proc. Nat. Acad. Sci.} \textbf{93},
  7436 (1996).

\bibitem[{Beenakker \& Staring(1992)}]{beenakker_theory_1992}
C. W.~J. Beenakker and A. A.~M. Staring, 
\newblock {Phys. Rev. B} \textbf{46}, 9667 (1992).

\bibitem[{Staring \emph{et~al.}(1993)}]{staring_coulomb-blockade_1993}
A. A.~M. Staring, L. W. Molenkamp, B.~W. Alphenaar, H. van Houten, O. J.~A. Buyk, M. A.~A. Mabesoone, C. W.~J. Beenakker and C.~T. Foxon,
\newblock {Europhys. Lett.} \textbf{22}, 57 (1993).

\bibitem[{Nakpathomkun \emph{et~al.}(2010)Nakpathomkun, Xu \&
  Linke}]{nakpathomkun_thermoelectric_2010}
N. Nakpathomkun, H.~Q. Xu and H. Linke, 
\newblock {Phys. Rev. B} \textbf{82}, 235428 (2010).

\bibitem[{Humphrey \emph{et~al.}(2002)Humphrey, Newbury, Taylor \&
  Linke}]{humphrey_reversible_2002}
T.~E. Humphrey, R. Newbury, R.~P. Taylor and H. Linke, 
\newblock {Phys. Rev. Lett.} \textbf{89}, 116801 (2002).

\bibitem[{Büttiker(1987)}]{buttiker_transport_1987}
M. B\"uttiker, 
\newblock {Z. Phys. B} \textbf{68}, 161 (1987).

\bibitem[{Esposito \emph{et~al.}(2009{\natexlab{a}})Esposito, Lindenberg and
  Van~den Broeck}]{esposito_thermoelectric_2009}
M. Esposito, K. Lindenberg and C. Van~den Broeck, 
\newblock {Europhys. Lett.} \textbf{85}, 60010 (2009{\natexlab{a}}).

\bibitem[{Sánchez \& Büttiker(2011)}]{sanchez_optimal_2011}
R. S\'anchez and M. B\"uttiker, 
\newblock {Phys. Rev. B} \textbf{83}, 085428 (2011).

\bibitem[{Sothmann \emph{et~al.}(2012)Sothmann, Sánchez, Jordan \&
  Büttiker}]{sothmann_rectification_2012}
B. Sothmann, R. S\'anchez, A.~N. Jordan and M. B\"uttiker, 
\newblock {Phys. Rev. B} \textbf{85}, 205301 (2012).

\bibitem[{Büttiker(1988)}]{buttiker_coherent_1988}
M. B\"uttiker, 
\newblock {{IBM} J. Res. Dev.} \textbf{32}, 63
  (1988).

\bibitem[{Edwards \emph{et~al.}(1993)Edwards, Niu and
  de~Lozanne}]{edwards_quantum-dot_1993}
H.~L. Edwards, Q. Niu and A.~L. de~Lozanne, 
\newblock {Appl. Phys. Lett.} \textbf{63}, 1815 (1993).

\bibitem[{Edwards \emph{et~al.}(1995)Edwards, Niu, Georgakis and
  de~Lozanne}]{edwards_cryogenic_1995}
H.~L. Edwards, Q. Niu, G.~A. Georgakis and A.~L. de~Lozanne,
\newblock {Phys. Rev. B} \textbf{52}, 5714 (1995).

\bibitem[{Prance \emph{et~al.}(2009)}]{prance_electronic_2009}
J.~R. Prance, C.~G. Smith, J.~P. Griffiths, S.~J. Chorley, D. Anderson, G. A.~C. Jones, I. Farrer and D.~A. Ritchie,
\newblock {Phys. Rev. Lett.} \textbf{102}, 146602 (2009).

\bibitem[{Van~den Broeck(2005)}]{van_den_broeck_thermodynamic_2005}
C. Van~den Broeck,
\newblock {Phys. Rev. Lett.} \textbf{95}, 190602 (2005).

\bibitem[{Esposito \emph{et~al.}(2009{\natexlab{b}})Esposito, Lindenberg and
  Van~den Broeck}]{esposito_universality_2009}
M. Esposito, K. Lindenberg and C. Van~den Broeck, 
\newblock {Phys. Rev. Lett.} \textbf{102}, 130602 (2009{\natexlab{b}}).

\bibitem[{Benenti \emph{et~al.}(2011)Benenti, Saito and
  Casati}]{benenti_thermodynamic_2011}
G. Benenti, K. Saito and G. Casati, 
\newblock {Phys. Rev. Lett.} \textbf{106}, 230602 (2011).

\bibitem[{Drexler \emph{et~al.}(1994)Drexler, Leonard, Hansen, Kotthaus and
  Petroff}]{drexler_spectroscopy_1994}
H. Drexler, D. Leonard, W. Hansen, J.~P. Kotthaus and P.~M. Petroff, 
\newblock {Phys. Rev. Lett.} \textbf{73}, 2252 (1994).

\bibitem[{Jung \emph{et~al.}(2005)}]{jung_shell_2005}
 M. Jung, T. Machida, K. Hirakawa, S. Komiyama, T. Nakaoka, S. Ishida and Y. Arakawa,
\newblock {Appl. Phys. Lett.} \textbf{87}, 203109 (2005).

\bibitem[{Scheibner \emph{et~al.}(2005)Scheibner, Buhmann, Reuter, Kiselev \&
  Molenkamp}]{scheibner_thermopower_2005}
R. Scheibner, H. Buhmann, D. Reuter, M.~N. Kiselev and L.~W. Molenkamp, 
\newblock {Phys. Rev. Lett.} \textbf{95}, 176602 (2005).

\bibitem[{Averin \emph{et~al.}(1997)Averin, Korotkov, Manninen \&
  Pekola}]{averin_resonant_1997}
D.~V. Averin, A.~N. Korotkov,  A.~J. Manninen and J.~P. Pekola, 
\newblock {Phys. Rev. Lett.} \textbf{78}, 4821 (1997).

\bibitem[{Park \emph{et~al.}(2004)Park, Tatebayashi \&
  Arakawa}]{park_formation_2004}
S.-K.Park, J. Tatebayashi and Y. Arakawa, 
\newblock {Appl. Phys. Lett.} \textbf{84}, 1877 (2004).



\end{thebibliography}

\end{document}